\documentclass{scrartcl}

\usepackage[utf8]{inputenc}

\usepackage{fixltx2e}

\usepackage{amsthm}
\newtheorem*{remark}{Remark}
\newtheoremstyle{named}{}{}{\itshape}{}{\bfseries}{.}{.5em}{\thmnote{#1 }#3}
\theoremstyle{named}

\usepackage{microtype}
\usepackage{amsmath}
\usepackage{mathtools}
\usepackage{booktabs}
\usepackage{tabularx}

\usepackage{pgfplots}
\pgfplotsset{compat=newest}

\usepackage{siunitx}

\newcommand\mytitle{On the conversion from OSA-UCS to CIEXYZ}
\newcommand\myauthor{Nico Schlömer}

\usepackage[
  pdfencoding=unicode,
  ]{hyperref}
\hypersetup{
  pdfauthor={\myauthor},
  pdftitle={\mytitle}
}

\usepackage[T1]{fontenc}
\usepackage{newtxtext}
\usepackage{newtxmath}

\usepackage{gensymb}

\usepackage{amsmath}

\usepackage{bm}

\title{\mytitle\footnote{The LaTeX sources as well as the source code for all
experiments in this article are available on \url{https://github.com/nschloe/colorio}}}
\author{\myauthor}

\begin{document}

\maketitle

\begin{abstract}
This article revisits Kobayasi's and Yosiki's algorithm for conversion of OSA-UCS into
  XYZ cooordinates. It corrects some mistakes on the involved functions and initial
  guesses and shows that that hundreds of thousands of coordinates can be converted in
  less than a second with full accuracy.
\end{abstract}

In 1974, MacAdam published the definition of the OSA-UCS color space~\cite{macadam} that
tries to adhere particularly well to experimentally measured color distances. It
combines work that had been going on since the late 1940s. One aspect of OSA-UCS is
that, while the conversion from CIEXYZ coordinates into OSA-UCS $Lgj$ coordinates is
straightforward, the conversion the other way around is not. In fact, there is no
conversion method that works solely in elementary functions. Apparently, this had not
been a design goal of OSA-UCS although is severely limits the usability of OSA-UCS.

In 2002, Kobayasi and Yosiki presented an algorithm for conversion from $Lgj$ to $XYZ$
coordinates that leverages Newton's method for solving nonlinear equation
systems~\cite{kobayasi}. Unfortunately, the article remains vague at important points
and also contains false assertions about the nature of the involved functions.

In 2013, Cao et al.\ compared Kobayasi's and Yosiki's approach with some other, more
complex methods based on artificial neural networks and found the latter to be
superior~\cite{cao}.

In the present note, the author aims to iron out the inaccuracies in Kobayasi's article
and improves the overall efficiency of the algorithm.

\section{The forward conversion}

The conversion from CIEXYZ coordinates to OSA-UCS $Lgj$ coordinates is defined as
follows:

\begin{itemize}
  \item Compute $x$, $y$ coordinates via
    \[
      x = \frac{X}{X + Y + Z},\quad y = \frac{Y}{X + Y + Z}.
    \]
  \item Compute $K$ and $Y_0$ as
    \begin{equation}\label{eq:KY0}
      \begin{split}
        K &= 4.4934 x^2 + 4.3034 y^2 - 4.276 x y - 1.3744 x - 2.5643 y + 1.8103,\\
        Y_0 &= Y K.
      \end{split}
    \end{equation}

  \item Compute $L'$ and $C$ as
    \begin{align}
        \label{eq:lc}
        L' &= 5.9 \left(\sqrt[3]{Y_0} - \frac{2}{3} + 0.042 \sqrt[3]{Y_0 - 30}\right)\\
        \nonumber
        C &= \frac{L'}{5.9 \left(\sqrt[3]{Y_0} - \frac{2}{3}\right)}.
    \end{align}
    (Note that $L'$ is $L$ in the original article~\cite{macadam}.)

  \item Compute RGB as
    \begin{equation}\label{eq:m}
      \begin{bmatrix}
        R\\
        G\\
        B
      \end{bmatrix}
      =
      M
      \begin{bmatrix}
        X\\
        Y\\
        Z
      \end{bmatrix}
      \quad\text{with}\quad
      M=\begin{bmatrix}
        +0.7990 & 0.4194 & -0.1648\\
        -0.4493 & 1.3265 & +0.0927\\
        -0.1149 & 0.3394 & +0.7170
      \end{bmatrix}.
    \end{equation}

  \item Compute $a$, $b$ as
    \begin{equation}\label{eq:ab}
      \begin{bmatrix}
        a\\
        b
      \end{bmatrix}
      =
      A
      \begin{bmatrix}
        \sqrt[3]{R}\\
        \sqrt[3]{G}\\
        \sqrt[3]{B}
      \end{bmatrix}
      \quad\text{with}\quad
      A = \begin{bmatrix}
        -13.7 & +17.7 & -4\\
        1.7 & +8 & -9.7
      \end{bmatrix}.
    \end{equation}

  \item Compute $L$, $g$, $j$ as
    \[
      L = \frac{L' - 14.3993}{\sqrt{2}},\quad g = Ca,\quad j = Cb.
    \]
\end{itemize}

\section{The backward conversion}

This section describes the conversion from the  $Lgj$ to the $XYZ$ coordinates.

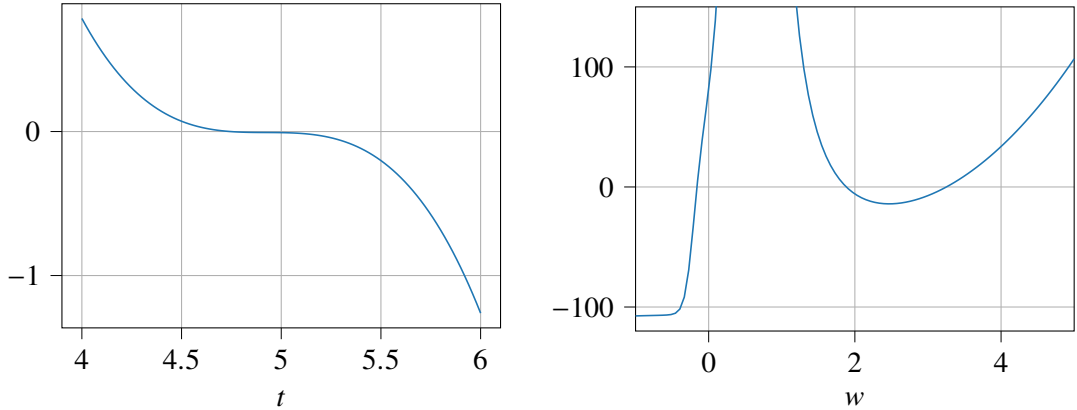
\begin{figure}
  \centering
  \hfill
  % This file was created by tikzplotlib v0.8.5.
\begin{tikzpicture}

\definecolor{color0}{rgb}{0.12156862745098,0.466666666666667,0.705882352941177}

\begin{axis}[
width=0.5\textwidth,
height=0.4\textwidth,
tick align=outside,
tick pos=left,
x grid style={white!69.01960784313725!black},
xmajorgrids,
xmin=3.9, xmax=6.1,
xlabel={$t$},
xtick style={color=black},
y grid style={white!69.01960784313725!black},
ymajorgrids,
ymin=-1.36340273736755, ymax=0.88755207389457,
ytick style={color=black}
]
\addplot [semithick, color0]
table {%
4 0.785235946109928
4.02 0.735087280230532
4.04 0.687106436386218
4.06 0.641245411020767
4.08 0.59745620057795
4.1 0.555690801501548
4.12 0.515901210235332
4.14 0.478039423223082
4.16 0.442057436908571
4.18 0.407907247735577
4.2 0.375540852147875
4.22 0.344910246589242
4.24 0.315967427503453
4.26 0.288664391334285
4.28 0.262953134525512
4.3 0.238785653520913
4.32 0.216113944764262
4.34 0.194890004699335
4.36 0.175065829769908
4.38 0.156593416419758
4.4 0.13942476109266
4.42 0.123511860232391
4.44 0.108806710282726
4.46 0.0952613076874412
4.48 0.0828276488903126
4.5 0.0714577303351169
4.52 0.0611035484656296
4.54 0.0517170997256262
4.56 0.0432503805588832
4.58 0.0356553874091769
4.6 0.028884116720283
4.62 0.0228885649359772
4.64 0.017620728500036
4.66 0.0130326038562349
4.68 0.00907618744835042
4.7 0.00570347572015815
4.72 0.00286646511543439
4.74 0.000517152077954889
4.76 -0.00139246694850414
4.78 -0.00291039552016687
4.8 -0.00408463719325717
4.82 -0.00496319552399913
4.84 -0.00559407406861669
4.86 -0.0060252763833339
4.88 -0.00630480602437471
4.9 -0.00648066654796317
4.92 -0.00660086151032324
4.94 -0.00671339446767894
4.96 -0.00686626897625427
4.98 -0.00710748859227323
5 -0.0074850568719598
5.02 -0.00804697737153799
5.04 -0.00884125364723183
5.06 -0.00991588925526531
5.08 -0.0113188877518624
5.1 -0.013098252693247
5.12 -0.0153019876356434
5.14 -0.0179780961352754
5.16 -0.021174581748367
5.18 -0.0249394480311421
5.2 -0.029320698539825
5.22 -0.0343663368306394
5.24 -0.0401243664598096
5.26 -0.0466427909835592
5.28 -0.0539696139581127
5.3 -0.0621528389396935
5.32 -0.0712404694845263
5.34 -0.0812805091488343
5.36 -0.0923209614888424
5.38 -0.104409830060774
5.4 -0.117595118420853
5.42 -0.131924830125303
5.44 -0.147446968730349
5.46 -0.164209537792216
5.48 -0.182260540867126
5.5 -0.201647981511303
5.52 -0.222419863280971
5.54 -0.244624189732356
5.56 -0.268308964421681
5.58 -0.293522190905168
5.6 -0.320311872739043
5.62 -0.34872601347953
5.64 -0.378812616682854
5.66 -0.410619685905235
5.68 -0.4441952247029
5.7 -0.479587236632074
5.72 -0.516843725248979
5.74 -0.55601269410984
5.76 -0.59714214677088
5.78 -0.640280086788325
5.8 -0.685474517718396
5.82 -0.73277344311732
5.84 -0.782224866541318
5.86 -0.833876791546617
5.88 -0.887777221689438
5.9 -0.943974160526009
5.92 -1.00251561161255
5.94 -1.06344957850529
5.96 -1.12682406476044
5.98 -1.19268707393425
6 -1.26108660958291
};
\end{axis}

\end{tikzpicture}
  \hfill
  % This file was created by tikzplotlib v0.8.5.
\begin{tikzpicture}

\definecolor{color0}{rgb}{0.12156862745098,0.466666666666667,0.705882352941177}

\begin{axis}[
width=0.5\textwidth,
height=0.4\textwidth,
tick align=outside,
tick pos=left,
x grid style={white!70!black},
xmajorgrids,
xmin=-1, xmax=5,
  xlabel={$w$},
xtick style={color=black},
y grid style={white!70!black},
ymajorgrids,
ymin=-120, ymax=150,
ytick style={color=black}
]
\addplot [semithick, color0]
table {%
-1 -107.413027646396
-0.939393939393939 -107.309786831062
-0.878787878787879 -107.215237122039
-0.818181818181818 -107.127190104587
-0.757575757575758 -107.042359529441
-0.696969696969697 -106.954298360616
-0.636363636363636 -106.847025918721
-0.575757575757576 -106.673796503393
-0.515151515151515 -106.283050417142
-0.454545454545455 -105.164709350817
-0.393939393939394 -101.697626835742
-0.333333333333333 -91.7446608040379
-0.272727272727273 -69.1567235939574
-0.212121212121212 -33.4644220071798
-0.151515151515151 5.1529978523228
-0.0909090909090908 38.1524502469182
-0.0303030303030303 66.9186705474708
0.0303030303030303 97.8541264114677
0.0909090909090908 138.357047423149
0.151515151515152 197.483926762583
% 0.212121212121212 289.632097951798
% 0.272727272727273 442.766820835499
% 0.333333333333333 720.341273215897
% 0.393939393939394 1294.7437109525
% 0.454545454545455 2781.23788859758
% 0.515151515151515 8852.923932832
% 0.575757575757576 140918.648042227
% 0.636363636363636 39329.2370995238
% 0.696969696969697 6307.75066265103
% 0.757575757575758 2481.55333720743
% 0.818181818181818 1314.67993372474
% 0.878787878787879 807.312619103627
% 0.939393939393939 540.585092146669
% 1 382.62864181619
% 1.06060606060606 281.115967891114
% 1.12121212121212 211.877730132073
1.18181818181818 162.475784128249
1.24242424242424 125.967621812498
1.3030303030303 98.2246601475964
1.36363636363636 76.6635115820041
1.42424242424242 59.5986674360327
1.48484848484848 45.8913206375379
1.54545454545455 34.748884678301
1.60606060606061 25.6055084607501
1.66666666666667 18.048190278627
1.72727272727273 11.7696062648161
1.78787878787879 6.53714074322924
1.84848484848485 2.17204573916972
1.90909090909091 -1.46489446641638
1.96969696969697 -4.48341990771192
2.03030303030303 -6.97097581197619
2.09090909090909 -8.99792296863166
2.15151515151515 -10.6213738820869
2.21212121212121 -11.8880550669845
2.27272727272727 -12.836469002897
2.33333333333333 -13.4985456088261
2.39393939393939 -13.9009169725394
2.45454545454545 -14.0659108181188
2.51515151515152 -14.0123317432564
2.57575757575758 -13.7560807176327
2.63636363636364 -13.310650174996
2.6969696969697 -12.6875225826085
2.75757575757576 -11.8964935122321
2.81818181818182 -10.9459352060274
2.87878787878788 -9.84301290556648
2.93939393939394 -8.59386342878862
3 -7.20374338230765
3.06060606060606 -5.6771528032405
3.12121212121212 -4.01793880526886
3.18181818181818 -2.22938286355019
3.24242424242424 -0.314274643424397
3.3030303030303 1.72502529216217
3.36363636363636 3.88653200963904
3.42424242424242 6.16859043355976
3.48484848484849 8.56983602610926
3.54545454545455 11.0891608914696
3.60606060606061 13.7256844599801
3.66666666666667 16.4787280530134
3.72727272727273 19.3477927478875
3.78787878787879 22.3325400585856
3.84848484848485 25.4327750269647
3.90909090909091 28.6484313839621
3.96969696969697 31.9795584937718
4.03030303030303 35.4263098382213
4.09090909090909 38.9889328353565
4.15151515151515 42.6677598169073
4.21212121212121 46.4632000149567
4.27272727272727 50.375732429673
4.33333333333333 54.405899468091
4.39393939393939 58.5543012592435
4.45454545454546 62.8215905639128
4.51515151515152 67.2084682082878
4.57575757575758 71.7156789801974
4.63636363636364 76.3440079346035
4.6969696969697 81.0942770619015
4.75757575757576 85.9673422784606
4.81818181818182 90.9640907039071
4.87878787878788 96.0854381940253
4.93939393939394 101.332327101924
5 106.705724243391
};
\end{axis}

\end{tikzpicture}
  \hfill
  \caption{Left: Graph of $f(t)$ \eqref{eq:f} for $L'=25$. Note that the root is not in
  the turning point, but close to it. This is because of the small second term in $f$.
  Right: Graph of the function $\phi$ for $L$, $g$, $j$ computed from $X=12$, $Y=67$,
  $Z=20$. The singularity is at $w\approx 0.59652046418$.  Note that the function as
  three roots only the largest of which is of interest.}\label{fig:singularity}
\end{figure}

Given $L$, we can first compute
\[
  L' = L \sqrt{2} + 14.3993.
\]
Equation~\eqref{eq:lc} gives the nonlinear relationship between $L'$ and $Y_0$ from
which we will retrieve $Y_0$. First set $t\coloneqq \sqrt[3]{Y_0}$ and consider
\begin{equation}\label{eq:f}
  0 = f(t) \coloneqq {\left(\frac{L'}{5.9} + \frac{2}{3} - t\right)}^3 - 0.042^3 (t^3 - 30).
\end{equation}
$f$ is a monotonically decreasing cubic polynomial (see figure~\ref{fig:singularity}).

Hence, it has exactly one root that can be found using the classical Cardano formula:

\begin{itemize}
  \item Expand $f(t) = at^3 + bt^2 + ct + d$ with
    \[
      \begin{split}
        &u = \frac{L'}{5.9} + \frac{2}{3},\quad v = 0.042^3,\\
        &a = -(v + 1),\quad  b = 3u,\quad  c = -3u^2, \quad d = u^3 + 30v.
      \end{split}
    \]

  \item Compute the depressed form $f(t)=a(x^3 + px + q)$:
    \[
      p = \frac{3ac - b^2}{3a^2},\quad q = \frac{2b^3 - 9abc + 27a^2d}{27a^3}.
    \]

  \item Compute the root as
    \[
      t = -\frac{b}{3a} + \sqrt[3]{
        -\frac{q}{2} + \sqrt{{\left(\frac{q}{2}\right)}^2 + {\left(\frac{p}{3}\right)}^3}
      }
      + \sqrt[3]{
        -\frac{q}{2} - \sqrt{{\left(\frac{q}{2}\right)}^2 + {\left(\frac{p}{3}\right)}^3}
      }.
    \]
    Note that the expression in the square root, ${\left(q/2\right)}^2 +
    {\left(p/3\right)}^3$ is always positive since, as argued above, $f$ has
    exactly one root.
\end{itemize}

\begin{remark}
  Kobayasi and Yosiki find the root of $f$ it using Newton's method.  A good initial
  guess here is $t = \frac{L'}{5.9} + \frac{2}{3}$ since the second term in $f(t)$,
  containing $0.042^3$, is usually small. Indeed it typically only takes around 10
  iterations to converge to machine precision.

  Cardano's method finds the root at once at the expense of computing one square root
  and two cube roots. This approach is found to be about 15 times faster.
\end{remark}

From here, one can compute
\begin{equation}\label{eq:gather}
  Y_0 = t^3,\quad
  C = \frac{L'}{5.9 \left(t - \frac{2}{3}\right)},\quad
  a = \frac{g}{C},\quad
  b = \frac{j}{C}.
\end{equation}
With $a$ and $b$ at hand, it is now possible via equation~\eqref{eq:ab} to pin down
$(\sqrt[3]{R}, \sqrt[3]{G}, \sqrt[3]{B})$ to only one degree of freedom, $w$.
The exact value of $w$ will be found by Newton iteration. The function $\phi(w)$
of which a root needs to be found is defined as follows.
\begin{quotation}
  Append the matrix $A$~\eqref{eq:ab} with a row such that the new $3\times3$-matrix
  $\tilde{A}$ is nonsingular and solve
  \[
    \begin{bmatrix}
      a\\
      b\\
      w
    \end{bmatrix}
    =
    \tilde{A}
    \begin{bmatrix}
      \sqrt[3]{R}\\
      \sqrt[3]{G}\\
      \sqrt[3]{B}
    \end{bmatrix}
  \]
  (Kobayasi, for instance, appends $[1, 0, 0]$ which corresponds to setting
  $w=\sqrt[3]{R}$.) Then compute the tentative $\tilde{X}$, $\tilde{Y}$, $\tilde{Z}$
  via~\eqref{eq:m} and further get the corresponding tentative $\tilde{Y}_0$
  from~\eqref{eq:KY0}. Then $\phi(w) = \tilde{Y}_0(w) - Y_0$.
\end{quotation}

If the difference between $\tilde{Y}_0(w)$ and $Y_0$ from~\eqref{eq:gather} is 0, the
correct $w$ has been found.  Kobayasi states the function $\phi$ is ``monotone
increasing, convex downward, and smooth''. Unfortunately, none of this is true (see
figure~\ref{fig:singularity}). In fact, the function has a singularity at $w$ chosen
such that the computed tentative $\tilde{X}$, $\tilde{Y}$, $\tilde{Z}$ sum up to 0 while
the individual values of $|\tilde{X}|, |\tilde{Y}|, |\tilde{Z}| > 0$. This happens if
the tentative $[R, G, B]$ is orthogonal on $[1,1,1] M^{-1}$.

Fortunately, it seems that the function is indeed convex to the right of the
singularity.  Newton's method will hence find the correct (largest) root if the initial
guess $w_0$ is chosen larger than the root. Since $w$ corresponds to $\sqrt[3]{R}$, it
is reasonable to chose $w_0$ to be the maximum possible value that $\sqrt[3]{R}$ can
take, namely that corresponding to $X=Y=100$, $Z=0$ (see~\eqref{eq:m}), $w_0=\sqrt[3]{79.9
+ 41.94}\approx 4.9575$.

\begin{figure}
  \centering
  \hfill
  % This file was created by tikzplotlib v0.8.5.
\begin{tikzpicture}
\small

\definecolor{color0}{rgb}{1,0.498039215686275,0.0549019607843137}
\definecolor{color1}{rgb}{0.12156862745098,0.466666666666667,0.705882352941177}
\definecolor{color2}{rgb}{0.83921568627451,0.152941176470588,0.156862745098039}
\definecolor{color3}{rgb}{0.580392156862745,0.403921568627451,0.741176470588235}
\definecolor{color4}{rgb}{0.172549019607843,0.627450980392157,0.172549019607843}

\begin{axis}[
width=0.5\textwidth,
height=0.4\textwidth,
legend cell align={left},
legend style={at={(0.03,0.97)}, anchor=north west, draw=white!80.0!black},
log basis x={10},
log basis y={10},
tick align=outside,
tick pos=left,
x grid style={white!69.01960784313725!black},
xmajorgrids,
xmin=0.466516495768404, xmax=8990687.44201965,
xmode=log,
xtick style={color=black},
y grid style={white!69.01960784313725!black},
ylabel={Runtime [s]},
ymajorgrids,
ymin=2.54437866616101e-07, ymax=22.9198831115375,
ymode=log,
ytick style={color=black},
xlabel={Size of the array}
]
\addplot [semithick, color1]
table {%
1 0.000471044
2 0.000493364
4 0.000493436
8 0.000496653
16 0.000569121
32 0.000455073
64 0.000551353
128 0.000530194
256 0.000632414
512 0.001044303
1024 0.001434569
2048 0.001990027
4096 0.003218893
8192 0.005218656
16384 0.013776633
32768 0.044884756
65536 0.089765479
131072 0.162647182
262144 0.408672322
524288 1.01199472
1048576 1.894782626
2097152 3.468689603
4194304 7.81700846
};
\addlegendentry{OSA-UCS}
\addplot [semithick, color2]
table {%
1 8.9266e-05
2 9.2652e-05
4 9.3239e-05
8 9.3739e-05
16 9.9035e-05
32 9.8968e-05
64 0.000105689
128 0.000118602
256 0.00014335
512 0.000189389
1024 0.000271935
2048 0.00042539
4096 0.000735595
8192 0.001374689
16384 0.002681221
32768 0.008891838
65536 0.02621592
131072 0.049708766
262144 0.095939109
524288 0.194949605
1048576 0.396554787
2097152 0.851998144
4194304 1.165202175
};
\addlegendentry{CIECAM02}
\addplot [semithick, color4]
table {%
1 1.9605e-05
2 2.0169e-05
4 2.0567e-05
8 2.0203e-05
16 2.262e-05
32 2.065e-05
64 2.195e-05
128 2.8066e-05
256 3.9437e-05
512 3.0937e-05
1024 3.8563e-05
2048 5.544e-05
4096 0.000155565
8192 0.00028523
16384 0.000265324
32768 0.000501102
65536 0.002711648
131072 0.007986971
262144 0.013183137
524288 0.025345891
1048576 0.067077266
2097152 0.08853523
4194304 0.170746303
};
\addlegendentry{CIELAB}
\end{axis}

\end{tikzpicture}
  \hfill
  % This file was created by tikzplotlib v0.8.5.
\begin{tikzpicture}

\definecolor{color0}{rgb}{0.12156862745098,0.466666666666667,0.705882352941177}
\definecolor{color1}{rgb}{1,0.498039215686275,0.0549019607843137}

\begin{axis}[
width=0.5\textwidth,
height=0.4\textwidth,
log basis x={10},
tick align=outside,
tick pos=left,
x grid style={white!69.01960784313725!black},
xmajorgrids,
xmin=0.466516495768404, xmax=8990687.44201965,
xmode=log,
xtick style={color=black},
y grid style={white!69.01960784313725!black},
ymajorgrids,
ymin=0, ymax=830,
ytick style={color=black},
xlabel={Size of the OSA-UCS array}
]
\addplot [semithick, color0]
table {%
1 787.066433566434
2 748.043062200957
4 631.69647696477
8 540.053088803089
16 430.216724738676
32 197.28831378892
64 135.024555079973
128 69.6830099941211
256 42.7417648170309
512 37.6977113374785
1024 25.8176709590747
2048 20.1362440723177
4096 14.446842284113
8192 11.6819040387925
16384 16.5864959757117
32768 29.5641483554035
65536 31.3942138250337
131072 24.4242208227254
262144 30.4130064747132
524288 37.476925997706
1048576 34.5039382592277
2097152 31.8295473370856
4194304 39.7625721082598
};
\end{axis}

\end{tikzpicture}
  \hfill
  \caption{Computation speed for arrays of $Lgj$ values measured with
  colorio~\cite{colorio}. Left: Comparison with CIELAB and CIECAM02.
  The conversion of several hundred thousand $Lgj$ values takes about 1 second. Right:
  Computation speed relative to the evaluation of the cubic root. For large arrays, the
  conversion to $XYZ$ is about as costly as the evaluation of 35 cubic roots.}\label{fig:speed}
\end{figure}
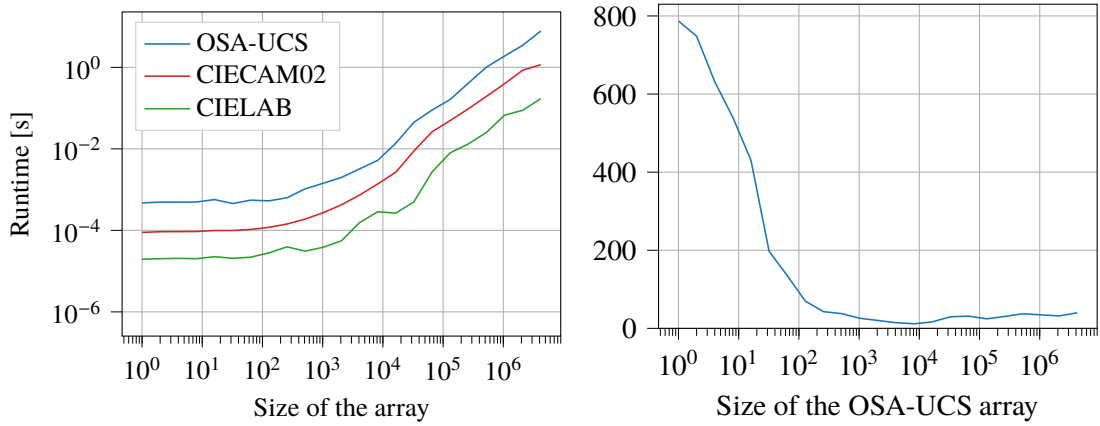

\begin{remark}
  Cao et al.~\cite{cao} found that the conversion to from $Lgj$ to $XYZ$ takes so long
  that alternative methods needs to be researched. They even find that the Newton
  iterations sometimes do not converge, or find the correct result only to few digits of
  accuracy.  The author cannot confirm these observations. The computation of hundreds
  of thousands of coordinates at once merely takes a second of computation time on a
  recent computer (figure~\ref{fig:speed}).

  To achieve this speed, it is important to vectorize all computation, i.e., not to
  perform the conversion for each $Lgj$-tuple individually one after another, but to
  perform all steps on the array. This also means to perform the Newton iteration on all
  tuples until the last one of them has converged successfully, even if some already
  converge in the first step. The redundant work inflicted by this approach is far
  outweighed by the advantages of vectorization.

  All code is published as open-source in colorio~\cite{colorio}.
\end{remark}

% \printbibliography{}
\bibliography{main}{}
\bibliographystyle{plain}

\end{document}